\documentclass[12pt]{article}
\usepackage[dvips]{graphicx}
\if@twoside \oddsidemargin 26pt
 \else \oddsidemargin 12pt\evensidemargin 12pt
\fi
\newcommand{\be}{\begin{eqnarray}}
\newcommand{\ee}{\end{eqnarray}}
\begin{document}
\title{Vortices and bags in $2+1$ dimensions}
\author{C.~D.~Fosco$^a$~\footnote{Electronic address:
    fosco@cab.cnea.gov.ar} and A.~Kovner$^b$~\footnote{Electronic
    address: a.kovner1@physics.ox.ac.uk}
  \\ \\
  {\normalsize\it $^a$Centro At\' omico Bariloche - Instituto Balseiro,}\\
  {\normalsize\it Comisi{\'o}n Nacional de Energ\'{\i}a At{\'o}mica}\\
  {\normalsize\it 8400 Bariloche, Argentina}\\ \\
  {\normalsize\it $^b$Department of Physics, Theoretical Physics}\\
  {\normalsize\it University of Oxford}\\
  {\normalsize\it 1 Keble Rd, Oxford OX1 3NP, UK}}
\maketitle
\begin{abstract}
\noindent
We consider the effect of the (heavy) fundamental quarks on the low
energy effective Lagrangian description of nonabelian gauge theories
in 2+1 dimensions. We show that in the presence of the fundamental
charges, the magnetic $Z_N$ symmetry becomes local. We construct the
effective Lagrangian representing this local symmetry in terms of
magnetic vortex fields, and discuss its physical consequences. We show
that the finite energy states described by this Lagrangian have
distinct bag-like structure. The point-like quarks are confined to the
region of space where the value of the vortex field is much smaller
than in the surrounding vacuum.
\end{abstract}
\bigskip
\newpage
\section{Introduction.}
The understanding of confinement and, more generally, of the principles
that govern the low energy dynamics in QCD is, perhaps, the most
interesting unanswered question in the theory of strong interactions.
The strongly coupled dynamics of the four dimensional non Abelian gauge
theories has so far proven to be forbiddingly complicated.  A fruitful
approach to cope with such situation can be to turn to simpler, more
tractable models, that nevertheless capture at least some of the
important features of the real theory.

 From this point of view, three dimensional non Abelian gauge theories
are of particular interest since they are, indeed, confining, and have
a strong coupling dynamics. On the other hand they can be continuously
deformed into the weak coupling region without encountering any phase
transition. The prototypical case is the $SU(2)$ gauge theory with an
adjoint Higgs field - the Georgi-Glashow model.  These weakly coupled
theories are confining, just like their strongly coupled counterparts.
Here, however, confinement can be studied using semiclassical methods,
as first done by Polyakov for the $SU(2)$ case~\cite{polyakov}, and
later on generalized to $SU(N)$ theories~\cite{sun}.

A related feature of these models is that their spectra has a large
gap between the light (light `photons') and heavy (heavy massive gauge
bosons) states. This makes the identification of the relevant low
energy degrees of freedom relatively easy, and also allows one to
construct the low energy effective Lagrangian explicitly. It turns out
that this effective Lagrangian exhibits the confinement phenomenon in
a very simple and straightforward way {\em on the classical level}. In
a nutshell, the natural degrees of freedom turn out to be magnetic
vortex fields first introduced by 'tHooft~\cite{thooft}, and the
confinement mechanism is realized as confinement of topological
solitons for those vortex fields. The structure of the effective
Lagrangian is robustly determined by the spontaneously broken magnetic
$Z_N$ symmetry~\cite{thooft}.  This approach has been motivated by
earlier work~\cite{early,kovner1}, it was advanced and formulated
in~\cite{kovner2}, and further clarified in~\cite{kovner3}. It has
also been argued in~\cite{kovner2}, that the same type of effective
Lagrangian remains valid in the strongly coupled limit of pure
gluodynamics. In this situation, the components of the vortex field
are naturally identified with the scalar and pseudoscalar glueballs,
which according to lattice results are indeed the two lowest states in
the spectrum of $2+1$ dimensional gluodynamics~\cite{teper}. The
confinement mechanism for the strongly coupled regime remains, in this
picture, essentially the same as for the weakly coupled phase.

All previous work dealt exclusively with theories having no matter
fields in fundamental representation.  The purpose of this note is to
address the question of what effect dynamical fundamental `quarks'
have on this low energy structure. That that effect exists and must be
nontrivial can be seen from one basic observation: in the presence of
fundamental charges, the vortex field (which is the effective low
energy degree of freedom) is no longer a local field. The effective
Lagrangian cannot, therefore, be a simple local scalar theory as
in~\cite{early,kovner1}.  We will show below that, in the presence of
fundamental matter, the magnetic $Z_N$ in the effective Lagrangian is
{\em gauged}, the value of the gauge coupling being inversely
proportional to the mass of the (lightest) dynamical quark.  We will
construct such an effective Lagrangian, which moreover in the limit of
the infinitely heavy quarks reduces to the effective theory
of~\cite{kovner1}.  The presence of fundamental quarks implies the
existence of a conserved baryon number charge. We will see how this
charge is represented in the low energy theory, and will study the
structure of finite energy baryon states.  Quite surprisingly, we find
that these states naturally have a bag-like structure. The value of
the vortex field in the region of space which contains the quarks, is
much lower than its value in the vacuum. The potential energy
associated with this difference in the expectation values can also be
naturally identified with the bag constant.

The plan of this paper is the following. In section~\ref{glue}, we
give a lightning review of the results of~\cite{kovner1}. We discuss
the magnetic $Z_N$ symmetry, its order parameter (the vortex
operator), and the structure of the effective Lagrangian in pure
gluodynamics. In section~\ref{fund} we explain the basic physics of
the changes in this structure in the presence of the fundamental
charges. In section~\ref{low} we show how to define the relevant low
energy degrees of freedom, and how to construct the effective
Lagrangian in this case. In section~\ref{baryon}, we study the
structure of the baryon states. Finally we conclude in
section~\ref{discus} with a short discussion.

\section{Pure glue.}\label{glue}
\subsection{The magnetic $Z_N$.}
Consider the pure SU(N) gauge theory in $2+1$ dimensions
\begin{equation}
L=-{1\over 4}{\rm Tr}F^2
\end{equation}
As was argued by 'tHooft in~\cite{thooft}, this theory possesses a
global $Z_N$ symmetry which is spontaneously broken in the vacuum.
The explicit realization of this symmetry was found in \cite{kovner3}.
The generator of the $Z_N$ group is (up to a multiplicative constant
factor) the fundamental Wilson loop along the spatial boundary of the
system
\begin{equation}
G=\lim_{C\to\infty}{\rm Tr}P\exp\{i\oint_C dx_iA^i(x)\} \;.
\end{equation}
In~\cite{kovner3}, the conservation of $G$ was proved taking the pure
gluodynamics limit of the Georgi-Glashow model. This, however, can
also be shown directly by calculating the commutator of $G$ with the
Yang-Mills Hamiltonian $H=1/2(E^2+B^2)$:
\begin{equation}
[G,H]=\lim_{C\to\infty}\oint_C dx_i
{\rm Tr} \left[ P E_i(x)\exp\{i\oint_{C(x,x)} dy_iA^i(y)\} \right]
\to_{C\to\infty}0 \;.
\label{commutator}
\end{equation}
Here, the integral in the exponential on the right hand side starts
and ends at the point of insertion of the electric field.  The
vanishing of the commutator follows from the fact that it only
involves electric fields at spatial infinity, and in a theory with a
finite mass gap those fields should vanish at infinity.

To dispel any doubt about the vanishing of this commutator we note
that the situation is completely analogous to the commutator of any
`conserved charge' which is defined as an integral of a local charge
density
\begin{equation}
Q=\lim_{C\to\infty} \int_{|x|\in C}d^2x\rho(x)\;.
\end{equation}
The commutator of such a charge with a Hamiltonian contains also a
surface term, since the charge density $\rho$ never commutes with the
Hamiltonian density but rather gives a total derivative in the commutator.  For a
conserved charge, due to the continuity equation, this surface term is
equal to the circulation of the spatial component of the current
\begin{equation}
[Q,H]=i\oint_C dx^ij_i\;.
\end{equation}
The vanishing of this term again is the consequence of the vanishing
of the physical fields at infinity in a theory with a mass gap. When
the charge is not conserved, the commutator contains also a bulk term
in addition to the surface contribution. It is the absence of the bulk
terms that is the distinctive property of a conserved charge.  The
same conclusion is reached if, rather than considering the generator
of the algebra, one considers the commutator of the group element for
either continuous or discreet symmetry groups.  We see, therefore, that
the commutator in (\ref{commutator}) indeed tells us that $G$ is a
conserved operator.

Next, a simple argument establishes that the magnetic $Z_N$ symmetry
is spontaneously broken in the vacuum if the Wilson loop $W_C$ has an
area law behaviour~\cite{kovner3}.  The argument goes like this. The
VEV of $W_C$ is the overlap of the vacuum state $|0>$ with the state
$|S>$ which is obtained from the vacuum by acting on it with $W_C$.
When acting on the vacuum state, $W_C$ performs the $Z_N$
transformation at all points within the area $S$ bounded by the loop.
If the vacuum wavefunction depends on the configuration of the $Z_N$
non invariant degrees of freedom (the state in question is not $Z_N$
invariant) the action of $W_C$ affects the state everywhere inside the
loop.  In the local theory with finite correlation length the overlap
between the two states approximately factorizes into the product of
the overlaps taken over the region of space of linear dimension of
order of the correlation length $l$
\begin{equation}
<0|S>=\Pi_x<0_x|S_x>
\label{fact}
\end{equation}
where the label $x$ is the coordinate of the point in the center of a
given small region of space. For $x$ outside the area $S$ the two
states $|0_x>$ and $|S_x>$ are identical and therefore the overlap is
unity. However for $x$ inside $S$ the states are different and the
overlap is therefore some number $e^{-\gamma}$ smaller than unity. The
number of such regions inside the area is obviously of order $S/l^2$
and thus
\begin{equation}
<W_C>=\exp\{-\gamma{S\over l^2}\}
\end{equation}
The VEV of $W_C$ then falls off as an area.  Conversely, if the vacuum
is $Z_N$ invariant, the wavefunction does not depend on the
configuration of the non invariant degrees of freedom. The action of
$W_C$ then alters the state only along the perimeter and $W_C$ has
the perimeter law behaviour in the unbroken phase.

Thus the area law behaviour of the fundamental Wilson loop is
tantamount to the breaking of the magnetic $Z_N$ symmetry in the
vacuum of pure gluodynamics.

\subsection{The vortex field.}
The only requirement that this argument presupposes is the existence
of {\bf local} degrees of freedom which are non invariant under $Z_N$,
or in other words the existence of a {\bf local} order parameter.
Such an operator can indeed be constructed explicitly.  This is the
magnetic vortex creation operator
$V$~\cite{early},\cite{kovner2},\cite{kovner3}.  The defining property
of $V$ is that it satisfies the following commutation relation with
the spatial fundamental Wilson loop $W$
\begin{equation}
V^\dagger(x)W(C)V(x)=e^{i{2\pi\over N} n(x,C)}W(C)
\end{equation}
where $n(x,C)$ is the linking number of the curve $C$ and the point
$x$, or in other words, $n=1$ if $x\in S$, and $n=0$ if $x\notin S$,
where $S$ is the region of the plane bounded by $C$. An explicit
representation for $V(x)$ is given by writing it as the operator that
performs a `singular gauge transformation'
\begin{equation}
V(x)=\exp\{{2i\over gN}\int dy_i \epsilon_{ij}{x_i-y_j\over (x-y)^2}
{\rm Tr}(YE_j(y))+ \Theta(x-y)J_0^Y(y)\} \;.
\label{v1}
\end{equation}
Here, the hypercharge generator $Y$ is defined as
\begin{equation}
Y={\rm diag} \left(1,1,\ldots,-(N-1)\right),
\end{equation}
the electric field is written in the matrix notation
$E_i=\lambda^aE^a_i$, with $\lambda^a$ denoting the $SU(N)$ generator
matrices in the fundamental representation, and $J_0^Y(x)$ is the
hypercharge density due to gluons $J^Y_0=ig{\rm Tr}Y[A_i,E_i]$.  The
function $\Theta(x)$ is the planar angle of the vector $x$.  Using
Gauss' law
\begin{equation}
\partial_iE^i=gA_i\times E_i \;,
\end{equation}
this operator (on the states that satisfy Gauss' law) may be
written in an alternative representation
\begin{equation}
V(x)=\exp\{{ 4\pi i\over gN} \int_C dy^i
\epsilon_{ij}\rm{Tr}(YE_i(y))\;.
\label{v2}
\end{equation}
The integration here is along the branch cut in the definition of the
planar angle $\theta(x)$, which is an infinite line that starts at the
point $x$ and goes to infinity.

There are two crucial properties of the definition eq.(\ref{v2}) that
insure that the operator $V$ is a local physical field:
\begin{enumerate}
\item{Gauge invariance.} The definition is not explicitly gauge
  invariant. Nevertheless, one can show~\cite{thooft},\cite{kovner3}
  that $V$ transforms a physical state into another physical state.
  That is to say that the matrix elements of $V$ between a physical
  and an unphysical state vanish.  By physical state we mean a state
  which satisfies the non Abelian Gauss' law. By virtue of this
  property the definition eq.(\ref{v2}) can be used as long as the
  matrix elements of $V$ (or any power of $V$ with insertions of an
  arbitrary number of gauge invariant operators) are calculated
  between physical states.
\item{Locality.} Again, the definition is not explicitly local, since
  it contains the integral of the electric field along an infinite
  line.  However in a theory that does not contain fundamental
  charges, $V$ in fact does not depend on the curve $C$, but only on
  its endpoint $x$.  Physically, this is due to the fact that the
  Bohm-Aharonov phase between $V$ and any charged state in the theory
  vanishes.  More formally, consider changing the position of the
  curve $C$ to $C'$.  This adds to the phase in the definition
  eq.(\ref{v2}) a contribution ${4\pi\over gN}\int_{S}d^2x {\rm
    Tr}\partial_iYE^i$, where $S$ is the area bounded by $C-C'$.  In
  our normalization, the hypercharge of gluons is $0$ or $\pm gN/2$
  and no particles with smaller value of the hypercharge are present
  in the theory.  Therefore the hypercharge within any closed area is
  an integer multiple of the gauge coupling $\int_{S}d^2x
  \partial_iE^i={gN\over 2}n$, and the extra phase factor is always
  unity.  One can also show directly~\cite{kovner1}, using canonical
  commutation relations, that $V(x)$ commutes with all the local gauge
  invariant operators $O(y)$, unless $y=x$, which formally establishes
  the locality of $V$.
\end{enumerate}
\subsection{The low energy effective theory.}
The universal properties of the low energy dynamics of the pure Yang
Mills theory are determined by the spontaneous breaking of the
magnetic $Z_N$ symmetry. The effective low energy Lagrangian is
constructed in terms of the vortex field $V$\footnote{For $N>4$ in
principle we have to include in this Lagrangian higher order
potential terms which stabilize the potential energy at large values
of $V$. Although we will not indicate them explicitly, it is assumed
in the following that they are indeed present whenever necessary.}
\begin{equation}
{\cal L}=\partial_\mu V^*\partial^\mu V -\lambda(V^*V-\mu^2)^2
-\zeta(V^N +V^{*N}) + \ldots
 \label{ldualgg}
\end{equation}
This Lagrangian can be derived in a controllable way in the weakly
coupled phase\cite{kovner1},\cite{kovner2}.  The coupling constants in
eq.(\ref{ldualgg}) are determined in the weakly coupled region from
perturbation theory and dilute monopole gas approximation.  It was
argued in \cite{kovner2} that the same low energy Lagrangian is also
valid in the strongly coupled regime (pure gluodynamics).

Several features of this Lagrangian are worth noting:
\begin{itemize}
\item First, $<V>\neq 0$, and therefore the $Z_N$ symmetry is spontaneously
broken.
\item Second, the field components of $V$ can be reinterpreted in terms of
the physical glueball fields. Using a polar decomposition for $V$
\begin{equation}
V=\rho e^{i\chi} \;,
\end{equation}
we find that the fluctuation field $\rho$ is the charge conjugation
even scalar, while the phase $\chi$ is the conjugation odd
pseudoscalar. Indeed, according to lattice results~\cite{teper},
the two lightest glueballs in the spectrum of pure gluodynamics are
the scalar $0^{++}$ and pseudoscalar $0^{--}$ states.  The
Lagrangian (\ref{ldualgg}), is thus the natural effective Lagrangian
that one would write to describe the dynamics of the lightest
particles. In this respect the identification of the glueball fields
with the modulus and the phase of $V$ constrains the possible terms in
the effective Lagrangian by imposing the $Z_N$ symmetry on possible
interaction terms.

\item Finally, this effective Lagrangian exhibits in a simple way confining
properties of the theory.  The colour charged `gluon' states (or
massive vector bosons in the weakly coupled regime) are represented in
the low energy theory by topological solitons carrying unit winding of
the field $V$.  If not for the $U(1)$ symmetry breaking term in the
potential in (\ref{ldualgg}), the lowest lying state in the
topologically nontrivial sector would have a rotationally invariant
hedgehog-like structure. Its energy would be logarithmically divergent
in the infrared, which corresponds to the logarithmic Coulomb
potential in 2+1 dimensions.  However since the symmetry of the
potential is $Z_N$ rather than $U(1)$, the energy of the hedgehog
configuration diverges quadratically with the volume. This is so since
the number of vacuum states is discreet and in the hedgehog
configuration the phase of the field $V$ is far from its vacuum value
everywhere in space. The configuration that actually minimizes the
energy in the one soliton sector breaks rotational invariance by
confining the winding of the phase of $V$ to some quasi
one-dimensional strip, like in figure 1.  It is clear that,
dynamically, the width of this strip is of the order of the inverse
glueball mass and the strip itself is nothing but the confining
electric flux string.  The energy of such an isolated soliton now
diverges {\it linearly\/} in the infrared, due to the finite energy
density per unit length of the string.
\end{itemize}
\begin{figure}[!]
\label{fig1}
\centering
\includegraphics{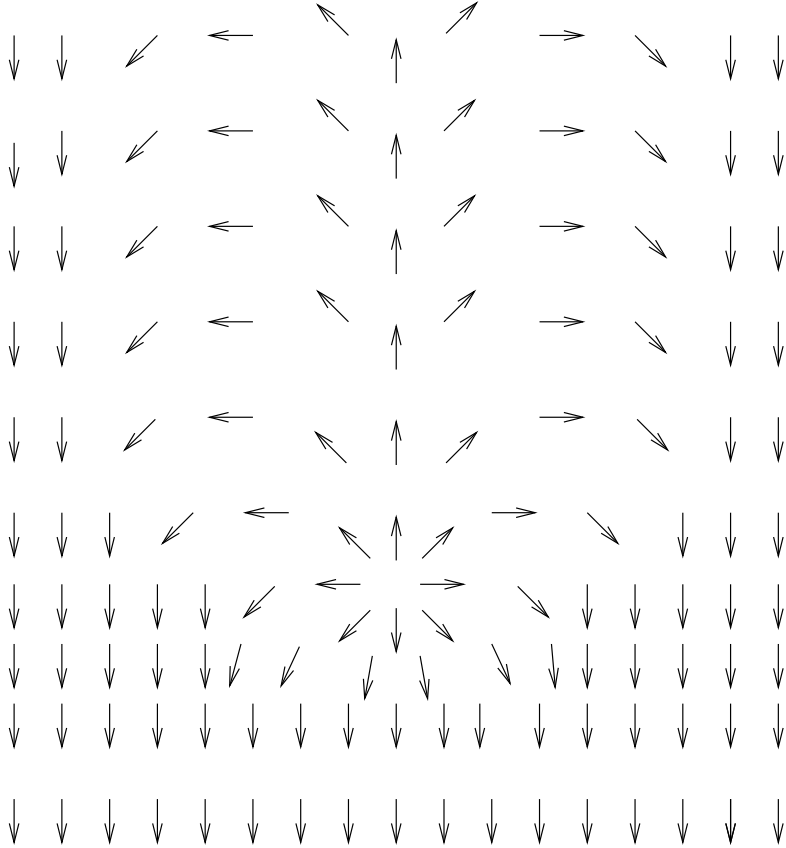}
\caption{The  string-like configuration of the field $V$
  in the state of unit charge in the presence of the symmetry breaking
  terms in the effective Lagrangian, for the $Z_2$ model.}
\end{figure}
The soliton we are discussing is a dynamical particle in the effective
theory - it thus represents a particle with an adjoint color charge,
and the string is the adjoint string. This string is of course not
absolutely stable since it can break into a soliton-antisoliton pair
when it is too long. The fundamental string also appears naturally in
the low energy description. It is the stable domain wall which
separates the regions of space with different expectation values of the
vortex field $V$ \cite{thooft,kovner2}, see figure 2.
\begin{figure}[!]
\label{fig2}
\centering
\includegraphics{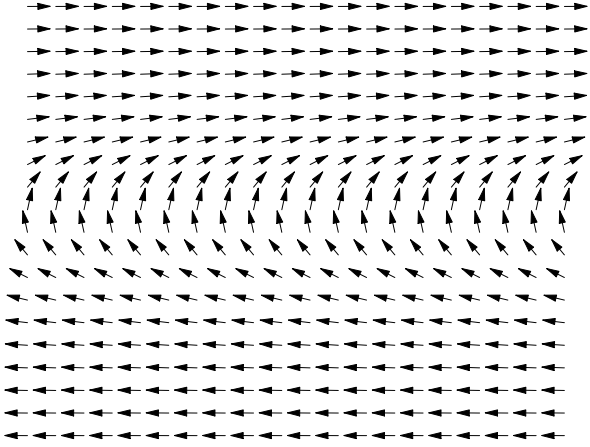}
\caption{The  domain wall representing the fundamental string in the
  effective theory.}
\end{figure}

\section{Magnetic $Z_N$ and fundamental quarks.}\label{fund}
In this section we wish to discuss the modifications introduced by the
presence of fundamental quarks in the structure just described. For
simplicity we will consider scalar quarks. The Lagrangian of the
theory we are interested in is
\begin{equation}
L=-{1\over 4}{\rm Tr}F^2+|D_\mu\Phi|^2-M^2\Phi^*\Phi
\label{lqcd}
\end{equation}
where the scalar field $\Phi^\alpha$ transforms according to the
fundamental representation of the $SU(N)$ color group.
\subsection{No local order parameter.}
First thing to note is that the fundamental Wilson loop still commutes
with the Hamiltonian. This is obvious, since the extra terms in the
Hamiltonian in the presence of the quarks do not involve electric
field operator, but only vector potential. The Wilson loop commutes
with the vector potential, and thus with the additional terms in the
Hamiltonian.  We thus conclude that the theory still has the magnetic
$Z_N$ symmetry.

This may seem somewhat surprising at first sight.  We have seen in the
previous section that spontaneous breaking of the magnetic $Z_N$
implies the area law for the Wilson loop, and conversely the perimeter
law of $W$ implies unbroken $Z_N$. In the theory with fundamental
charges the Wilson loop is known to have perimeter law due to breaking
of the confining string at any finite value of the fundamental mass
$M$.  We might then conclude that the magnetic $Z_N$ is restored at
any, arbitrarily large value of $M$ but is broken at
$M\to\infty$. The common lore is that the $Z_N$ breaking phase
transition for $N>2$ is first order, and this then implies a
discontinuous behaviour of the theory in the infinite mass limit. This
of course is completely counterintuitive and in fact plain wrong.  The
caveat in this line of reasoning is the following. The relation
between the behaviour of the Wilson loop and the mode of the
realization of the magnetic symmetry hinges crucially on the existence
of a local order parameter of the magnetic $Z_N$. In the absence of
such an order parameter it is not true that the Wilson loop locally changes
the quantum state inside the loop and this invalidates the whole
argument. In particular in the absence of a local order parameter, the
$Z_N$ symmetry can be spontaneously broken but the Wilson loop can
have a perimeter law.

In fact it is easy to see that for any finite $M$ the magnetic $Z_N$
does not have a local order parameter. The only candidate for such an
order parameter is the vortex operator $V(x)$ defined in
eq.(\ref{v2}), since it has to be local also relative to the purely
gluonic operators\footnote{One can of course multiply $V(x)$ by any
  explicitly local gauge invariant operator. Such modifications are
  however irrelevant as far as the locality properties of $V$ are
  concerned, and we will not consider them in the following.}.
However in the presence of fundamental quarks the operator $V(x)$ is
not local anymore.  To see this consider the
dependence of $V_C(x)$ on the curve $C$ which enters its definition.
As before the operators $V_C$ and $V_{C'}$ are related by
\begin{equation}
V_C(x)=V_{C'}(x)\exp\{{4\pi i\over gN}\int_{S}d^2x {\rm Tr}\partial_iYE^i\}
\end{equation}
where $S$ is the area bounded by $C-C'$. As before, due to the Gauss'
law the integral in the exponential is equal to the total hypercharge
in the area $S$.  However the hypercharge of fundamental quarks has
eigenvalues $\pm g/2$. The extra phase factor is therefore not unity
anymore but can rather take values $\exp\{2\pi i/N\}$ depending on the
state and the choice of the contour $C$.

The status of the magnetic $Z_N$ is thus quite different in a theory
with fundamental quarks - it does not have a local order parameter.
Nevertheless it is clear that at least as long as the mass $M$ is
large $M/g^2>1$, the relevant degrees of freedom for the effective
infrared dynamics should still be the vortices $V$ and the main factor
which determines their dynamics should still be the magnetic $Z_N$. At
large $M$ the dependence of $V$ on the curve $C$ is weak, since the
probability of the vacuum fluctuations which involve fundamental
charges is small.  The probability of appearance of a virtual $q\bar
q$ pair separated by a distance $l$ is suppressed by the exponential
factor $\exp\{-Ml\}$.  The typical distance scale for the "glueball"
physics is $1/g^2$.  Thus at these distances such fluctuations are
unimportant and should not affect much the dynamics.  Things are
different however if one is interested also in the baryonic sector of
the theory. The baryons are necessarily heavy and in order to be able
to discuss their structure we must understand the main dynamical
effects also at shorter distances.

Therefore our aim now is to understand
what is the main effect of the non locality discussed above on the
dynamics of magnetic vortices.
\subsection{$Z_N$ as a local symmetry.}
The situation we have just described - a symmetry without a local
order parameter - is not exceptional in quantum field theory. This is
precisely the property of the global part of any Abelian gauge group.
Consider for example quantum electrodynamics. The global electric
charge is of course a physical gauge invariant charge with the
corresponding gauge invariant local charge density.
\begin{equation}
Q=\int d^2x\rho
\end{equation}
Nevertheless there is no local operator that carries this charge. This
is a direct consequence of the Gauss' law
\begin{equation}
\partial_iE_i=g\rho
\end{equation}
Any physical, gauge invariant operator that carries $Q$ must also
carry the long range electric field, which can not fall off faster
than a power of the distance.  The gauge invariant QED Lagrangian is
written in terms of "local" charged fields $\phi$. But appearances are
deceptive: these fields are not gauge invariant, and therefore not
physical. A gauge invariant charged field can be constructed from
$\phi$ by multiplying it by a phase factor
\begin{equation}
\phi_{phys}(x)=\phi\exp\{ig\int d^2y e_i(x-y)A_i(y)\}
\end{equation}
with the $c$-number field $e_i$ satisfying
\begin{equation}
\partial_ie_i=\delta^2(x)
\end{equation}
For any $e_i$ satisfying this condition the operator $\phi_{phys}$ is
gauge invariant, and therefore physical. It is however necessarily
nonlocal. Different choices of $e_i$ define different gauge invariant
operators and correspond to different gauge fixings. Thus for
$e_i(x)=x_i/x^2$ the field $\phi_{phys}$ is the field $\phi$ in the
Coulomb gauge, while $e_i(x)=\delta_{i1}\delta(x_2)\theta(x_1)$
corresponds to the axial gauge $A_1=0$, and so on.  Different
definitions of $\phi_{phys}$ differ from each other by a phase factor,
which is precisely the gauge ambiguity of the original field $\phi$.

The $U(1)$ gauge group is the most natural Abelian gauge symmetry to
consider in continuum field theory. One can however also consider
discrete groups like $Z_N$\cite{zn}.  In this case again no local
operator that carries the global $Z_N$ charge exists.  The various
gauge invariant charged operators are nonlocal and differ from each
other by a local $Z_N$ - valued phase. These different operators again
correspond to different gauge fixings of the local $Z_N$ group.

This is precisely the structure that emerged in our discussion in the
earlier part of this section.  We have

1. Global magnetic $Z_N$ symmetry generated by the fundamental Wilson
loop.

2. The set of nonlocal vortex operators $V_C(x)$, which all carry the
$Z_N$ charge and differ from each other by a $Z_N$ valued phase
factors.

It is very suggestive therefore to think about $V_C(x)$ as of
different gauge fixed versions of a field charged under local $Z_N$.
This leads us to expect that the low energy theory we are after should
be a $Z_N$ gauge theory of the magnetic vortex field $V$.

In fact coming back to the discussion in the beginning of this
section, we see that from this vantage point it is obvious why the
Wilson loop has a perimeter law, even if the global $Z_N$ is broken
spontaneously. The action of a Wilson loop of a finite size inside the
contour is a gauge transformation. Thus in physical terms locally
inside the contour the new state is {\it the same\/} as the old one and
so the overlap between the two locally is unity. The only nontrivial
contributions to the overlap come from the region close to the contour,
thus giving the perimeter law.

In the next section we shall construct the effective gauge theory and
discuss its relation to the original QCD Lagrangian.

\section{Gauging the Wilson loop.}\label{low}

Although it is possible to  give a more `analytic' derivation of the effective
Lagrangian, we prefer to write it down directly guided by the previous
discussion. We will then explain the physical meaning of the various
fields that appear in it.

\subsection{The Lagrangian.}
The easiest way to construct a $Z_N$ gauge theory in the continuum is
to consider a $U(1)$ gauge theory with the Higgs field of charge $N$
which has a large expectation value~\cite{zn}.  Consider therefore the
following Lagrangian:
\begin{eqnarray}
\label{efflfund} L&=&-{1\over 4e^2}f_{\mu\nu}^2+|(\partial_\mu-i{1\over
N}b_\mu)V|^2+ |(\partial_\mu-ib_\mu)U|^2 \nonumber\\
&-&\lambda(V^*V-\mu^2)^2 - \xi(V^NU^* +V^{*N}U) -
\tilde\lambda(U^*U-u^2)^2 \;.
\end{eqnarray}
Here $f_{\mu\nu}=\partial_\mu b_\nu-\partial_\nu b_\mu$. We take the
parameters such that, $\tilde\lambda>>\lambda$ and $u^2>>\mu^2$.

The `Higgs' field $U$ has a large expectation value and breaks the
$U(1)$ gauge symmetry down to its $Z_N$ subgroup
$V(x)\to\exp\{i{2\pi n(x)\over N}\}V(x)$. It should be noted that this gauge
subgroup does not correspond to `singular' gauge transformations,
since partial derivatives commute on the gauge parameter, despite the
fact that there are $\delta$-like singularities.

Below the scale determined by the expectation value $u$, the field $U$
is practically frozen and its fluctuations are unimportant. In this
regime the model indeed describes the locally $Z_N$ invariant theory.
The global part of the gauge group is our $Z_N$ magnetic symmetry
generated by the Wilson loop.  The larger $U(1)$ gauge structure at
this point is just an auxiliary trick which enables us to write down a
discrete gauge theory in continuous notations.  We will however see
later that it does in fact has a real physical meaning of its own and
arises naturally in the effective theory.

Let us first discuss how this Lagrangian reduces to the effective
Lagrangian of the pure Yang-Mills theory eq.(\ref{ldualgg}) in the
limit of the large quark mass.  In this limit not only the field $U$
must decouple, but also the gauge interactions of the field $V$ must
vanish.  There is another consistency requirement. In the limit of
zero gauge coupling $e^2\to 0$ the gauge $U(1)$ symmetry
becomes global $U(1)$ and is broken due to non vanishing expectation
value of $U$. The spectrum therefore contains a massless Goldstone
boson. This Goldstone boson is of course the longitudinal component of
the gauge field $b_\mu$. There is however no such massless particle in
the pure gluodynamics nor in the effective Lagrangian
eq.(\ref{ldualgg}).  This means that the couplings in
eq.(\ref{efflfund}) should depend on the quark mass in such a way that
the vector particles remains heavy for any finite $M$ and its mass
goes to zero very sharply only in the limit when it is completely
decoupled.  In terms of the Goldstone boson couplings it means that
$f_\pi$ must be larger than any scale relevant to the dynamics of the
vortex field $V$.  All these conditions can be met by choosing for
example
\begin{equation}
e^2={y\over M}, \ \ \ \ \ \ \ u^2=xM,\ \ \ \ \ \ \ \ \ \xi={\zeta\over u}
\label{largem}
\end{equation}
With this choice the mass of the vector boson $m^2=e^2u^2$ stays
finite as $M\to \infty$ and can be arbitrarily large. The
Goldstone boson in the decoupling limit has an infinite $f_\pi$ and is
completely decoupled just as the "invisible axion".

\subsection{The symmetries and the fields.}

We now want to understand how the basic fields present in the
effective Lagrangian arise in the fundamental theory eq.(\ref{lqcd}).
We start with considering the symmetries.

The Lagrangian eq.(\ref{efflfund}) has two global $U(1)$ symmetries.
One is the global part of the local $U(1)$ with the conserved current
$ej^T_\mu=\partial_\nu f_{\nu\mu}$.  Note that due to the direct
coupling between the $V$ and $U$ fields the $U(1)$ rotations of $V$
and $U$ separately are not symmetries.  The other current, conserved
by virtue of the homogeneous Maxwell equation is $\tilde
f_\mu=\epsilon_{\mu\nu\lambda}f_{\nu\lambda}$.  The two charges have
quite different nature. The second one, the dual magnetic flux
\begin{equation}
\Phi_D=\int d^2 x \tilde f_0
\end{equation}
has a local order parameter. It can be constructed in a way similar to
the vortex field in QED \cite{early},\cite{kovner1}. The global $U(1)$
gauge charge, on the other hand does not have a local order parameter,
as discussed earlier.

Both these conserved currents should also exist in the fundamental
theory. It is fairly straightforward to identify them.  The QCD
Lagrangian has one obvious global $U(1)$ charge - the baryon number.
This charge has local order parameters - gauge invariant baryon fields
of QCD, and is therefore identified with the dual magnetic flux
\begin{equation}
{1\over 2\pi}\tilde f_\mu=J^B_\mu,\ \ \ \ \ \ Q_B=\Phi_D
\label{f}
\end{equation}

The second charge can be expressed in terms of the spatial current
components of the first one
\begin{equation}
Q^T=\int d^2xj^T_0=\int d^2x \partial_i\Big[{1\over e^2}\epsilon_{ij}\tilde f_j\Big]
\label{QT}
\end{equation}
It is thus the vorticity associated with the baryon number current.

The vortex operator $V$ which appears in eq.(\ref{efflfund}) is not
gauge invariant and is only physical after complete gauge fixing of
the $U(1)$ gauge group. After such a gauge fixing (which 
amounts to multiplying $V$ by an operator valued phase) the Gauss law
requires that on the physical states the physical operator $V$ carry
the charge $Q_T$.  Due to the identification eqs.(\ref{f},\ref{QT})
this leads to a surprising conclusion that any physical operator $V$
in the effective theory creates a vortex of the original baryon number
current.

This somewhat unexpected conclusion is in fact quite natural for an
eigenoperator of the magnetic $Z_N$. Consider the vortex operator $V$
defined by eq.(\ref{v2}). In QCD just like in pure gluodynamics, it
has an alternative representation of an operator of the singular gauge
transformation of the form eq.(\ref{v1})
\begin{equation}
V_C(x)=\exp\{{2i\over gN}\int d^y \epsilon_{ij}{x_i-y_j\over (x-y)^2}{\rm Tr}(YE_j(y))+
\theta(x-y)J_0^Y(y)\}
\label{v3}
\end{equation}
The only difference is that now $J_0^Y$ is the hypercharge operator
due to both, gluons and fundamental quarks
\begin{equation}
J^Y_0=ig\Big[{\rm Tr}Y[A_i, E_i]+Y_{\alpha\beta}(\Phi^*_\alpha\Pi_\beta-
\Phi_\alpha\Pi^*_\beta)\Big]
\end{equation}
Consider the action of this operator on a quark field $\Phi$. The
transformed quark field $\Phi'=V^\dagger_C\Phi V_C$ is a gauge
transform of $\Phi$ everywhere except along the branch cut of the
function $\theta$. Across this cut the phase of $\Phi'$ is
discontinuous - it jumps by $2\pi/N$ for all color components of
$\Phi'$.  Due to this discontinuity the baryon current - the global
$U(1)$ current of $\Phi$ - does not vanish at points along the cut.
Calculating explicitly the action of $V$ on the baryon number current
we find
\begin{equation}
V^\dagger_CJ^B_i(x)V_C=iV^\dagger_C(\Phi^*\partial_i\Phi-\partial_i\Phi^*\Phi)V_C=
J^B_i(x)+{2\pi\over N}n_i^C(x)\delta(x\in C)\Phi^*\Phi(x)
\end{equation}
where $n_i^C(x)$ is a unit vector normal to the branch cut $C$ at the
point $x$.

It is natural to define the local vorticity associated with the baryon
number as\footnote{For a one component field $\Phi$ the vorticity
  defined in this way is precisely the circulation of the phase of
  $\Phi$.}
\begin{equation}
\rho_T=i\epsilon_{ij}\partial_i[{(\Phi^*\partial_j\Phi-\partial_j\Phi^*\Phi)\over
\Phi^*\Phi}]
\label{vort}
\end{equation}
The vortex operator $V$ therefore creates a vortex of baryon number
current with fractional vorticity $2\pi/N$
\begin{equation}
V^\dagger_C(x)\rho_T(y)V_C(x)=\rho_T(y)+{2\pi\over N}\delta^2(x-y)
\end{equation}

The operator $U$ due to the Gauss' law also carries baryon vorticity.
Since its gauge coupling is $N$ times the coupling of $V$, it creates
one unit of vorticity.

In fact this simple exercise also helps us to identify the value of
the gauge coupling constant $e^2$ in the effective theory.  Comparing
eq.(\ref{vort}) with eq.(\ref{QT}) we find
\begin{equation}
e^2\propto \Phi^*\Phi
\label{e2}
\end{equation}
The same relation is obtained by comparing the current algebra in the
fundamental and the effective theories.  The commutator of the baryon
charge density with the baryon current density in the fundamental
theory is
\begin{equation}
[J^B_0(x),J^B_i(y)]=i\Phi^*\Phi\partial_i\delta^2(x-y)
\end{equation}
In the effective theory using the canonical commutators that follow
from eq.(\ref{efflfund}) and the identification eq.(\ref{f}) we find
\begin{equation}
[J^B_0(x),J^B_i(y)]=i{e^2\over 4\pi^2}\partial_i\delta^2(x-y)
\end{equation}
Again we deduce eq.(\ref{e2}).  The operator on the right hand side of
this equation in the effective theory is indeed a constant. Recall,
that our effective theory should be valid at long distances. In this
regime in the leading order in the derivative expansion the operator
$\Phi^*\Phi$ should be approximated by its expectation value.  Taking
into account fluctuations of $\Phi^*\Phi$ is tantamount to including
higher derivative terms in the effective Lagrangian
eq.(\ref{efflfund}).  At this, higher order in derivative expansion
the gauge coupling constant would become a dynamical field. While this
is perfectly legitimate, it is certainly beyond our present framework.

When taking the expectation value of $\Phi^*\Phi$ we should remember
that it has to be calculated with the cutoff $\Lambda$ appropriate for the
effective theory.  This cutoff must be above the characteristic scale
of the pure gluodynamics (determined by the string tension) but below
the heavy quark mass.  With this in mind we get
\begin{equation}
<\Phi^*\Phi>_\Lambda=\int_0^{\Lambda} {d^2p\over 8\pi^2}{1\over (p^2+M^2)^{1/2}}\propto
{\Lambda^2\over M}
\end{equation}
So that finally
\begin{equation}
e^2\propto{\Lambda^2\over M}
\end{equation}
which is consistent with the expected scaling in the large mass limit
eq.(\ref{largem}).

\section{The baryon and the bag.}\label{baryon}
\subsection{The baryon.}
Having understood the origin of the fields and the symmetries in the
effective theory, we would like to see how it encodes the qualitative
features of the low energy QCD physics.  Since we are considering a
heavy quark theory, below the fundamental mass scale the spectrum
should be the same as in pure gluodynamics.  We have already seen that,
in the infinite mass limit, the effective theory reduces to that of
(\ref{ldualgg}). Indeed, even at finite but large $M$, this is the case
at low energies. Formally, this can be seen as follows.  Since the VEV
of the field $U$ is large, we can impose the unitary gauge condition on
it. In this unitary gauge the phase of $U$ disappears. The modulus of
$U$ is very heavy, and so is the vector field $b_\mu$. Thus at low
energies we recover the effective theory of pure glue sector.  There is
however a set of configurations, on which the unitary gauge can not be
imposed.  Those are configurations in which $U$ vanishes at some points
in space.  Indeed it is these configurations that are important for the
baryonic sector.  Recall that $U$ is a vortex of baryon number current.
Thus one expects that the baryon charge is associated with the vortex
configuration of the field $U$. In the core of the vortex the field $U$
of course has to vanish, and so the unitary gauge is not admissible.
Thus in the baryon sector we can not think of $U$ as frozen at its
expectation value and instead have to treat it as a dynamical field.

That the baryon does indeed carry vorticity of the field $U$, can be
seen by the following simple argument.  The baryon number is
represented in the effective theory by the dual magnetic flux. The
baryonic state must therefore be the dual magnetic vortex.  Such a
vortex in a nonsingular gauge has a vector potential of the form
$b_i=\epsilon_{ij}{x_j\over x^2}$.  To have a finite energy it must be
accompanied by the winding of the phases of both $V$ and $U$.  In fact
since $U$ carries $N$ times the charge of $V$, the only states that
are allowed energetically are those that carry $N$ vortices of $U$.

This is natural in view of the `dual' relation between the effective
and the fundamental theories. The field $U$ is the `vortex' dual to
the fundamental quark.  We thus expect that its elementary vortex
would represent the fundamental quark itself, and so finite energy
states must contain $N$ such elementary vortices.  A single vortex
must be confined.  In the same sense, $V$ is `dual' to the adjoint
gluon field.  The elementary vortex of $V$ is then, in a sense, the
`constituent' gluon.  In fact, such single gluon should not exist as a
finite energy state either, and we expect it also to be confined. The
single vortex of $V$ should therefore bind either with the anti-vortex
of the same type, or with $N$ vortices of $U$.  The former type of
state is a glueball, and exists in the pure glue
theory~\cite{kovner2}, while the latter type is a baryon.

Interestingly enough, this line of reasoning leads us to expect that
the baryon must have a bag-like structure. Namely, the quarks are
bound to the $V$ field vortex. Inside this vortex, the value of $V$ is
small - in fact it exactly vanishes in the middle, and than rises
quite slowly (relative to the scale of $1/M$) towards the edges.
Recall that $V=0$ corresponds to a non confining state~\cite{kovner2}.
The quarks are therefore sitting in the `perturbative' region of space
- where there are no confining forces.  Only when they separate far
from each other - into the region with non vanishing $V$, the linear
potential pulls them inside again.

Let us look at this more carefully. Consider for simplicity the $Z_2$
symmetric case $N=2$.  The baryon is the dual vortex with dual
magnetic flux $2\pi/e$.  Far from the vortex core,
the field configuration is pure gauge, with the phases of $V$ and $U$
following the vector potential:
\begin{equation}
b_i=\epsilon_{ij}{x_j\over x^2},\;\;\; V(x)=ve^{i \alpha(x)},\;\;\;
U(x)=ue^{2 i \alpha(x)} \;.
\end{equation}
The parameters of the model are such that the field $V$ is much
lighter than both $U$ and $b_i$. Thus the size of the vortex core of
$V$ is large - of the order of the inverse glueball mass.
The two $U$-vortices which have a very small size core, sit inside
this core.
Since the length associated with the dual magnetic field is much
smaller than the core size of the $V$ - vortex, the dual flux is
concentrated on the $U$ - vortices. From the low energy
point of view, the picture is that two point-like magnetic vortices sit
inside a soft core of a $V$ field vortex.  The field configuration looks
roughly as depicted on figure 3.

The magnetic flux is concentrated in the points $A$ and $B$. The phase
of the field $U$ follows the variation of the vector potential very
closely. The phase of $V$ is also trying to do that, but it can not
quite follow it all the way, since on the line between the two
vortices it would have to be discontinuous. The most important energy
contribution (apart form the core energy of small vortices) comes
therefore from the vicinity of this line. The phase of $V$ obviously
has to interpolate across this line between the values $0$ and $\pi$.
Since the modulus of $V$ is not extremely rigid, it will be smaller
along this line than in the immediate neighborhood.  Both, the
variation of the modulus and the phase of $V$ along the line
connecting the small vortices contribute to the energy which is
clearly linear in the distance $|A-B|$.

\begin{figure}[!]
\label{fig3}
\centering
\includegraphics{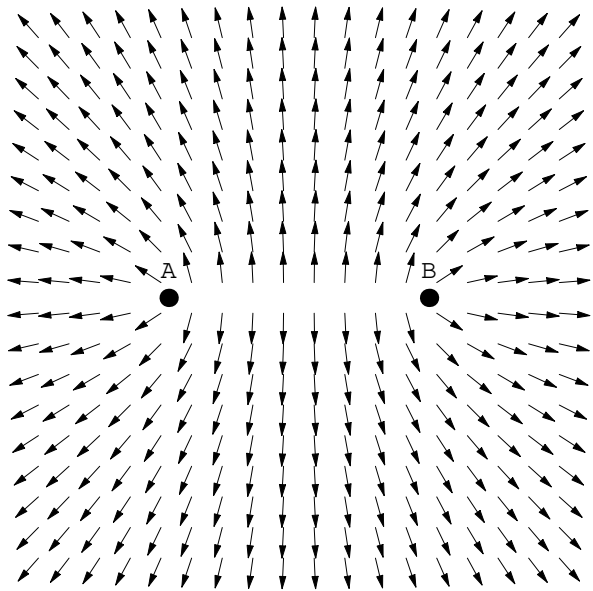}
\caption{The `baryon' configuration for $Z_2$. Arrows represent the
  direction (phase) of $V$, and the big dots at $A$ and $B$ correspond
  to the positions of the $U$ field vortices.}
\end{figure}

\subsection{The string and the bag.}
To study the structure of the baryon more quantitatively let us fix
the gauge such that
$$
 U(x)\,=\,u\, \exp [ i \, \theta (x) \,]
$$
\begin{equation}
 \label{eq:uconf}
\theta (x) \;=\; \arctan (\frac{y}{x-a}) + \arctan(\frac{y}{x+a}) \;.
\end{equation}
This is a valid gauge choice for the configuration with
two vortices of unit winding at the positions $(a,0)$
and $(-a,0)$.

Due to the conditions on the parameters of our model, the vector field
$b_\mu$ follows the phase of $U$ in the whole space. Thus
\begin{equation}
  \label{eq:bconf}
b_0=0,\ \ \ \ \ b_j = - \left[\epsilon_{jk}\frac{(x - x^{(l)})_k}{|x - x^{(l)}|^2}
\,+\, \epsilon_{jk}\frac{(x - x^{(r)})_k}{|x - x^{(r)}|^2} \right]
\end{equation}
where: $x_1 \equiv x$, $x_2 \equiv y$, and $l$, $r$ are the positions
of the left and right vortices, respectively: $x^{(l)} = (-a,0)$,
$x^{(r)} = (a,0)$.

There are two interesting limiting situations. The first is when the
distance $a$ is larger than the dynamical distance scales
$(\lambda\mu^2)^{-1/2}$ and $(\xi u^2)^{-1/2}$ which determine the
masses of the glueballs. The second interesting situation is the
reverse, that is when the two vortices are sitting well inside the
glueball correlation length. Let us look at them in turn.

When the distance between the vortices is large we expect the
potential between them to be linear with the string tension calculated
in the theory without the dynamical $U$ field.
We will study the interquark potential in our low energy theory in the
semiclassical approximation.

To find the minimal energy configuration, we have to solve the
classical equations of motion for the field $V$ at a fixed
configuration of $U$ and $b_i$ given by
eqs.(\ref{eq:uconf},\ref{eq:bconf}). Let us concentrate on the points
which are close to the $x$ - axis, with $|x|<<a$. The main contribution
to the energy comes from this region of space. In this region the
vector potential $b_i$ vanishes, and the phase of the field $U$ is
zero. Thus the equations of motion for the field $V$ are the same as in
the pure gluodynamics. Also as long as we stick to this region, the
configuration of $V$ is translationally invariant in the $x$ direction.
What determines the energy then are the boundary conditions on the
field $V$. In this configuration clearly the phase of the field $V$ is
$\pi/2$ far above the $x$ axis and $-\pi/2$ far below the axis, fig.3.
Thus both, the equations and the boundary conditions are precisely the
same as for the domain wall separating the two degenerate vacua in the
effective theory of pure gluodynamics ($e^2=0,\ \
\tilde\lambda\to\infty$). There is an extra contribution to the
energy that comes from the region of space close to the points $A$ and
$B$. But this energy does not depend on $a$ and is subleading for large
$a$. The rest of the space does not contribute to the energy, since the
field configuration there is pure gauge.

Thus as expected the energy in this regime is $E=a\sigma$, where
$\sigma$ is the domain wall tension (fundamental string tension)
calculated in pure gluodynamics~\cite{kovner2}.

It is perhaps more interesting to consider the opposite situation, that
is when the distance between the quarks is smaller than the glueball
correlation length. This is the regime in which we do not expect to see
any stringy structure. Instead we can ask whether the lowest energy
configuration has any resemblance to a bag. To study this question we
take the limit $a\to 0$. In this case clearly the phase of the
field $V$ will follow the phase of $U$ in the whole space
\begin{equation}
  \label{eq:imag1}
  (\frac{V}{V^*})^2 \;=\; \frac{U}{U^*} \;.
\end{equation}
Given this condition, only the variation of the radial component of
$V$ has to be determined. Since the problem has rotational symmetry,
the equation of motion for the modulus $\rho$
becomes
\begin{equation}
  \label{eq:eqrho1}
 -\frac{d^2 \rho}{d r^2}\,-\, 2 \eta  \rho
 + 2 \lambda \rho^3\;=\;0\;.
\end{equation}
with  $\eta \;=\; \lambda \mu^2 - \xi u$, which is assumed to be
positive throughout this paper. The relevant boundary condition in this
case is that at infinity $\rho$ approaches its vacuum value
$\rho_{r\to\infty}\to v$ while in the vortex core it
vanishes $\rho(0)=0$. With these boundary conditions, eq.
(\ref{eq:eqrho1}) has the familiar form of the $\varphi^4$ static kink
equation in $1+1$ dimensions with the solution
 \begin{equation}
  \label{eq:rhor}
  \rho (r) \;=\; v \;
  \frac{1 \,-\, e^{-2 \sqrt{\eta} r}}{1 \,+\, e^{-2 \sqrt{\eta} r}} \;.
\end{equation}
The energy of this solution is
\begin{equation}
E[V]_{a=0} \;=\; \lambda \int_0^\infty \, 2 \pi dr r \,
(v^4 - \rho^4) \;=\; \frac{\pi^3}{12} \, \frac{\eta}{\lambda}
\;=\; \frac{\pi^3}{12} \, ( \mu^2 - \frac{\xi u}{\lambda} )\;.
\end{equation}

The picture is thus just as described in the previous subsection. Since
the two quark state is accompanied by the vortex of the field $V$, the
quarks effectively `dig a hole' in the vacuum. In their immediate
vicinity the modulus $\rho$ vanishes, and therefore there is a `bag' of
the nonconfining state. The radius of this bag is given by the mass of
the scalar glueball $2 \sqrt{\eta}$.

It is interesting to note that, although for large separation $a$ the
energy of the string gets contributions from both, the scalar and the
pseudoscalar glueballs (the modulus and the phase of $V$), the `bag
constant' is determined solely by the scalar glueball. For small $a$
the phase of $V$ is not excited and only $\rho$ deviates from the
vacuum state inside the `bag'. This is consistent with the common lore
that the inside of the bag is distinguished from the outside by the
value of the $F^2$ condensate. In fact since $\rho$ has vacuum quantum
numbers and interpolates in our effective theory the scalar glueball,
it is naturally associated with the operator $F^2$ which has a large
overlap with the scalar glueball in QCD.

\section{Discussion}\label{discus}
In this note we have considered the effect of heavy fundamental quarks
on the effective low energy theory description of nonabelian gauge
theories in 2+1 dimensions. We found that the status of the magnetic
$Z_N$ symmetry changes. It becomes a local symmetry with the `gauge
coupling' inversely proportional to the quark mass. Thus the low energy
theory becomes a discrete gauge theory.

We have also studied the structure of the baryon in the framework of
the effective theory. The picture that emerges is very reminiscent to
the `bag'. The quarks in the baryon sit in the middle of the
$V$-vortex, where the energy density differs from the vacuum energy
density. The `bag constant' is equal to the difference of these two
energy densities, and parametrically in the effective theory is given
by $\eta v^2$, where $\eta$ is the mass of the scalar glueball and $v$
is the expectation value of the vortex radial field $\rho$.

An important thing to note is that the bag we are talking about here
arises in a very different situation than in the usual bag
model~\cite{bagmodel}. There the bag describes the structure of the
baryon containing {\it light\/} quarks. The radius of the bag in this
situation is determined by the balance of the vacuum pressure and the
pressure due to the free motion of the light quarks inside, and in fact
depends on the quark wave function. In our case the inside of the bag
contains heavy quarks. Their kinetic energy is small, and we have
treated them here as static. The radius of the bag thus is determined
purely by the dynamics of the scalar glueball field and is not
sensitive to the state of the heavy quarks. This is true for low lying
excitations for which the radius of the quark state is smaller than the
inverse glueball mass. When these two scales are comparable presumably
the quark pressure will also be important and will play a role in the
energy balance. Thus in this intermediate regime we expect the
$V$-vortex to be similar to the bag in the usual bag model. For states
of even larger size the potential between quarks is linear with the
fundamental string tension. The bag picture should therefore go
smoothly into the string picture. It would be interesting to study
these questions further in the framework of the effective theory
discussed in this note.

Lastly we note that although we have so far considered scalar quarks, 
extending this discussion to spin $1/2$ quarks is fairly
straightforward. It is well known \cite{chern} that in the Abelian
Higgs model the spin of the vortex is controlled by the coefficient of
the Chern-Simons term of the vector field. Thus adding the Chern
Simons term for the vector field $b_\mu$
\begin{equation}
\delta L= \frac{\kappa}{2}\epsilon^{\mu\nu\lambda}b_\mu\partial_\nu b_\lambda \;,
\end{equation}
with $\kappa = 1/(2 \pi)$, will endow the vortex of the $U$-field with a
half integer spin\footnote{The spin will reside on the $U$-vortex and
  not the $V$-vortex, since in the dynamics of our effective model the
  vector potential $b_\mu$ follows the phase of the field $U$.}.  Our
effective model with this extra term is relevant to the description of
QCD with one flavour of heavy spinor quarks.  Just like the fermion
mass term in 2+1 dimensions, the Chern Simons term breaks the parity
symmetry in the baryonic sector of the theory.  The extra term does
not affect the classical analysis we have preformed here, but is
clearly relevant for determination of the spin of the baryon as well
as the spectrum of the excitations.

\section*{Acknowledgements}
This work was part of a collaboration supported by a 
Fundaci{\'o}n Antorchas - The British Council
grant. C.~D.~F.~is a member of
CONICET,
 and is partially supported by ANPCyT (PICT 97/1040) and
Fundaci{\'o}n
 Antorchas grants.  A.~K.~is supported by PPARC.


\begin{thebibliography}{99}
\bibitem{polyakov}A.~M.~Polyakov, {\it Nucl.\ Phys.} {\it B120}, 429
(1977).
\bibitem{sun}S.~R.~Das and S.~R.~Wadia, {\it Phys.\ Lett.} {\bf B106}, 386
(1981), Erratum-ibid. {\bf B108} 435 (1982);
 N.~J.~Snyderman {\it Nucl.\ Phys.} {\bf B218}, 381 (1983).
\bibitem{thooft} G.~t'Hooft, {\it Nucl.\ Phys.} {\bf B138}, 1 (1978).
\bibitem{early}S.~Samuel, {\it Nucl.\ Phys.} {\bf B154}, 62,(1979); A.
  Kovner, B. Rosenstein and D. Eliezer, {\it Mod.\ Phys.\ Lett.} {\bf
    A5}, 2661 (1990); {\it Nucl.\ Phys.} {\bf B350}, 325 (1991); A.
  Kovner and B. Rosenstein, {\it Phys.\ Lett.} {\bf B266}, 443 (1991).
\bibitem{kovner1} A.~Kovner and B.~Rosenstein, {\it Int.\ J.\ Mod.\ Phys.}
{\bf A7}, 7419 (1992).
\bibitem{kovner2} A.~Kovner and B.~Rosenstein, {\it J.\ High\ Energy\ Phys.}
{\bf 003} 9809 (1998).
\bibitem{kovner3}C.~P.~Korthals-Altes and A.~Kovner, `Magnetic Z(N) symmetry in hot QCD
and the spatial Wilson loop,' hep-ph/0004052.
\bibitem{teper} M.~J.~Teper, {\it Phys.\ Rev.} {\bf D59} 014512 (1999).
\bibitem{zn} L. Krauss and F. Wilczek {\it Phys.\ Rev.\ Lett.} {\bf 62}, 1221 (1989);
 J. Preskill and L. Krauss {\it Nucl.\ Phys.} {\bf B341} 50 (1990).
\bibitem{bagmodel}A.~Chodos, R.~L.~Jaffe, K.~Johnson and C.~B.~Thorn,
{\it Phys.\ Rev.}  {\bf D10} (1974) 2599;
 A.~Chodos, R.~L.~Jaffe, K.~Johnson, C.~B.~Thorn and V.~F.~Weisskopf, {Phys.\
 Rev.} {\bf D9} (1974) 3471.
\bibitem{chern}See, for example: A. Lerda, {\em Anyons}, Lecture Notes in Physics,
  Springer-Verlag, Berlin, 1992.
\end{thebibliography}
\end{document}